\documentclass[prl,twocolumn,notitlepage,amsfonts,amssymb,amsmath,floatfix,tightenlines,superscriptaddress]{revtex4-2}

\usepackage{amsfonts,amssymb}
\usepackage{amsmath,times}
\usepackage{graphicx}
\usepackage[colorlinks=true,linkcolor=blue,anchorcolor=red,citecolor=blue,urlcolor=blue]{hyperref}
\usepackage{color}
\usepackage{cleveref}
\usepackage[lofdepth,lotdepth,caption=false]{subfig}

\begin{document}

\title{Anomalous Superfluid Density in Pair-Density-Wave Superconductors}
\author{Ke Wang }
\email{kewang07@uchicago.edu}
\affiliation{Department of Physics and James Franck Institute, University of Chicago, Chicago, Illinois 60637, USA}
\affiliation{Kadanoff Center for Theoretical Physics, University of Chicago, Chicago, Illinois 60637, USA}

\author{ Qijin Chen}
\affiliation{Hefei National Research Center for Physical Sciences at the Microscale and School of Physical Sciences, University of Science and Technology of China,  Hefei, Anhui 230026, China}
\affiliation{Shanghai Research Center for Quantum Science and CAS Center for Excellence in Quantum Information and Quantum Physics, University of  Science and Technology of China, Shanghai 201315, China}
\affiliation{Hefei National Laboratory,  Hefei 230088, China}
\author{Rufus Boyack}
\affiliation{Department of Physics and Astronomy, Dartmouth College, Hanover, New Hampshire 03755, USA}
\author{K. Levin}
\affiliation{Department of Physics and James Franck Institute, University of Chicago, Chicago, Illinois 60637, USA}

\date{\today}

\begin{abstract}
 Pair-density-wave (PDW) states are a long-sought-after phase of quantum
materials, with the potential to unravel the mysteries of high-$T_c$
cuprates and other strongly correlated superconductors. Yet, surprisingly, a key signature of stable superconductivity, namely the positivity of the superfluid density, $n_s(T)$, has not yet been demonstrated. Here, we address this central issue by calculating $n_s(T)$ for a generic model two-dimensional PDW superconductor. We uncover a surprisingly large region of intrinsic instability, associated with negative $n_s(T)$, revealing that a significant portion of the parameter space thought to be physical cannot support a pure PDW order. In the remaining stable regime, we predict two striking and observable fingerprints: a small longitudinal superfluid response and an unusual temperature dependence for $n_s(T)$. These generally model-independent, as well as experimentally relevant findings suggest that the fragility of the superfluid density poses a significant problem for the formation of stable, finite temperature PDW superconductivity.   
\end{abstract}
 
 \maketitle

\section{Introduction} 
The search for unconventional superconductors beyond the standard BCS paradigm\cite{PhysRev.108.1175} 
is a central driver of modern condensed matter physics
\cite{PhysRevLett.60.2677,Kong2025}. An intriguing phase is the Pair-Density-Wave (PDW) superconductor\cite{Agterberg2020,Soto2017,Wang2018,Huang2022,Wu1,Wu2,WuSteven2025,Chen2023,Rosales2024,Kong2025,papaj2025,WangKekule2025}, a state where Cooper pairs carry finite momentum, leading to a spatially oscillating order parameter. While such phases are thought to be observed in 
quantum materials,
there remains an essential aspect which is largely unexplored: namely, the behavior of
the superfluid density.
The superfluid density is a fundamental thermodynamic quantity that directly links the microscopic pairing of electrons to the macroscopic, measurable properties of the superconducting state.
Most importantly, its positivity is required for superfluid stability
~\cite{Pao2006,Gubankova2006,Chen2006}.
The absence of such a demonstration is a critical gap in the understanding of these exotic states.

\begin{figure}
    \centering
    \includegraphics[width=3.1in]{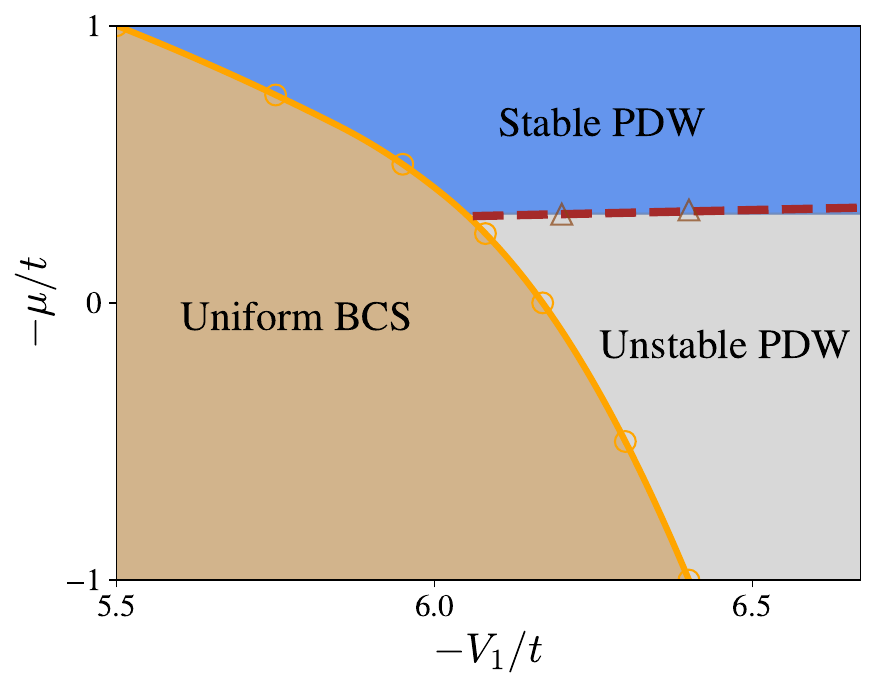}
 \caption{Zero-temperature phase diagram as a function of interaction strength ($V_1$) and chemical potential ($\mu$). The PDW state is stable only in a limited region (blue), separated from the uniform BCS phase by a first-order transition (solid line). A large portion of the phase diagram is an unstable regime (gray) characterized by a negative longitudinal superfluid density ($n_s^{xx} < 0$). There is a boundary (brown dashed) seperating stable and unstable PDW and determined by zero superfluid density.
 This finding guides the experimental search for stable PDW phases by identifying the strong-pairing, relatively low-density quadrant as the most promising regime. }
    \label{phase_diagram}
\end{figure}

Among experimentally prominent PDW states
is the unidirectional PDW, a class characterized by a time-reversal and inversion symmetric order parameter, $\Delta(\mathbf{r}) \sim \cos\left(\mathbf{Q}\cdot\mathbf{r}\right)$. Its defining feature is a large Cooper pair momentum, $\mathbf{Q}$, that is driven by electronic correlations and is comparable to a reciprocal lattice vector~\cite{Agterberg2020}. This phase has reportedly been observed in a diverse set of quantum materials, including high-$T_c$ cuprates~\cite{Du2020,Dai2018,Wang2018,Norman2018}, Fe pnictides~\cite{Zhao2023,papaj2025}, graphene-based systems~\cite{Venderley2019,Slagle2020,Yoshida2012,Chou2025,Han2022}, and transition metal dichalcogenides~\cite{Shaffer2023}. These superconductors are highly unusual, featuring an atypical excitation spectrum with gapless ``Fermi arcs''~\cite{Lee2014} and a high degree of entanglement~\cite{Brian2010PRL}.  
Experimental signatures of PDW order, such as modulated gaps\cite{Du2020,Zhao2023} have been reported.
Crucially, this state must be distinguished from the Fulde-Ferrell state~\cite{Fulde1964}, which emerges in a magnetic field that breaks time-reversal symmetry and is known to be unstable~\cite{Yin2014,Boyack2017}.

In this work, we microscopically investigate the superfluid density, $n_s(T)$, of 2D unidirectional PDW 
superconductors.
Beyond its critical role in determining stability, the superfluid density is crucial, for
its temperature dependence serves as a sensitive diagnostic, directly probing the unconventional excitation spectrum expected in these phases. Additionally, the superfluid density, which at low temperatures, is effectively
the  phase stiffness~\cite{PRL1977K,Prokof2000}, governs  
the rich melting phenomenology~\cite{Berg2009,Barci2011,Agterberg2008}.

Our investigation reveals two essential mechanisms behind the behavior of $n_s(T)$. The first mechanism which
stems from the finite center-of-mass momentum, $\mathbf{Q}$, of the Cooper pairs is associated with
destructive interference effects in superfluid transport. These severely suppress the magnitude of the superfluid density 
and, for unidirectional states, induce an extreme anisotropy. As a result
the positivity condition $n_s > 0$ is only met at sufficiently small $|\mathbf{Q}|$.
Also important is our experimental prediction of unconventional temperature dependences.
PDW order will generally induce a Van Hove singularity near the Fermi energy
in the fermionic excitation spectrum. This feature, related to the underlying Bogoliubov Fermi surface, 
leads to unconventional temperature dependences. The transverse component of the superfluid density {\it increases}, following
a $T^2$ power law at low temperatures ($n_s(T)-n_s(0) \propto + T^2$).

\vskip2mm

\section{Results}

\subsection{Lattice Model and PDW states} Our Hamiltonian is a
tight-binding model on a square lattice with nearest-neighbor
interactions:
\begin{equation}
 \hat{H} = \sum_{\sigma,\mathbf{k}} \xi_{\mathbf{k}}
c^\dagger_{\mathbf{k}\sigma} c_{\mathbf{k}\sigma} +
\sum_{\mathbf{k},\mathbf{k}',\mathbf{q}} V(\mathbf{q})
c^\dagger_{\mathbf{k}+\mathbf{q}\uparrow} c^\dagger_{\mathbf{k}’-\mathbf{q}
\downarrow} c_{\mathbf{k}' \downarrow} c_{\mathbf{k} \uparrow}.
\end{equation}
Here, $\sigma = \uparrow,\downarrow$ represents the spin, while the
fermionic dispersion is $\xi_{\mathbf{k}} = -2t(\cos k_x + \cos k_y) +
4t' \cos k_x \cos k_y - \mu$ with $t'/t=0.3$ and lattice constant set
to $a=1$. We use the momentum-space potential $V(\mathbf{q}) = 2V_1 [ \cos q_x + \cos q_y ]$ with $V_1<0$, which represents a nearest-neighbor attractive interaction consistent with experimental observations~\cite{Chen2021,PhysRevLett.127.197003} in related quantum materials. We restrict our analysis to a chemical potential, $\mu$, far from the band bottom so that our system is well away from the BEC limit. This supports the applicability of a mean-field,
intermediate coupling approach as
the interaction strength we consider remains smaller than the total electronic bandwidth.
 
 When $V_1$ is attractive, electrons are 
paired which will lead ultimately to a superconducting phase at low
$T$. For parameter regimes shown in Fig.~\ref{phase_diagram}, the energetically favored ground states are either a uniform superconductor
which is not of interest here, or a unidirectional PDW\cite{Loder2010,PhysRevB.96.224503}. We provide a discussion of possible alternative many-body phases in the Methods Section ("Possible Phases"). The unidirectional PDW is characterized by an order parameter, $\Delta_l(\mathbf{q})$, that is finite only at the ordering vectors $\mathbf{q} = \pm \mathbf{Q}$ and is defined by:
\begin{eqnarray}
    \Delta_l(\mathbf{q}) =  V_1 \sum_{\mathbf{p}} \varphi_l(\mathbf{p}) \langle c_{\mathbf{p}+\mathbf{q}/2 \uparrow} c_{-\mathbf{p}+\mathbf{q}/2 \downarrow} \rangle. 
\end{eqnarray}
Here $\varphi_l = \cos p_x + l \cos p_y$, with $l = \pm1$ representing $s$-
and $d$-wave orderings\footnote{Other symmetry channels are absent due to the reflection symmetry.}. 
The pairing momentum $\mathbf{Q}$ will be determined self-consistently.
Note that the presence of $\pm \mathbf{Q}$ breaks 4-fold rotational symmetry. As a consequence, the mixing of \( s \)- and \( d \)-wave order parameters is unavoidable\footnote{In the PDW phase plotted in Fig.~\ref{fig2}, \( \Delta_s({\bf Q}) / \Delta_d({\bf Q}) \simeq 0.4 \).}. Also see Fig.~\ref{FS0}.

 To solve for PDW states self-consistently, we employ the Gor'kov Green's function formalism~\cite{Loder2010}. While the full non-perturbative treatment (provided in Methods Section:``Gor'kov Equations") is complex, to leading order we can neglect correlations between the two condensates at $+\mathbf{Q}$ and $-\mathbf{Q}$. This simplification allows us to work with a reduced set of propagators: the normal Green's function, $ G(\tau, \mathbf{k}) = -T_\tau \langle c_{\mathbf{k}\sigma}(\tau) c^\dagger_{\mathbf{k}\sigma}(0) \rangle $, and two anomalous propagators, $ F_m(\tau, \mathbf{k}) = -T_\tau \langle c_{\mathbf{k} \uparrow}(\tau) c_{-\mathbf{k} + m \mathbf{Q} \downarrow}(0) \rangle $ with $m = \pm 1$, which describe pairing with momentum $\pm\mathbf{Q}$. They are coupled through the simplified Gor'kov equations:
\begin{eqnarray}
  G^{-1}(k) &=& G_0^{-1}(k) - \sum_{m} G_0^{-1}(-k + m\mathbf{Q}) \Delta^2(\mathbf{k}, m \mathbf{Q}), \nonumber\\
  F_m(k) &=& \Delta(\mathbf{k}, m \mathbf{Q}) G_0^{-1}(-k + m \mathbf{Q}) G(k).\label{4}
\end{eqnarray}
Here, $k = (i k_0, \mathbf{k})$ is the energy-momentum vector, $G_0$ is the free fermion propagator, and $ \Delta(\mathbf{k}, m \mathbf{Q}) = \sum_l \varphi_l(\mathbf{k}) \Delta_l(m \mathbf{Q}) $ is the order parameter weighted by symmetry factors.

The self-consistent PDW solution is found by simultaneously solving the gap equation, $\Delta_l(m{\bf Q}) = V_1 \sum_k \varphi_l(\mathbf{k}-m\mathbf{Q}/2) F_m(k)$~\cite{Chen2007,Loder2010}, and minimizing the ground state energy $\langle\hat{H}\rangle$ to find the optimal pairing momentum $\mathbf{Q}$. A stable PDW state with a global energy minimum at finite $\mathbf{Q}$ only exists when both $s$- and $d$-wave pairing components are allowed to mix~\footnote{If we restrict ourselves to a pure \(d\)-wave solution, no global minimum in \(\langle\hat{H}\rangle\) as a function of \(Q\) is found.}. The elementary excitations of this self-consistent state are Bogoliubov quasiparticles, whose dispersion, $E_\alpha(\mathbf{k})$, is given by the poles of the Gor'kov Green's functions~\footnote{A similar three-band analysis was presented in Ref.~\onlinecite{Chen2007}.}. This allows the propagators to be written in their spectral representation: $G = \sum_\alpha g_\alpha({\bf k}) [i k_0 - E_\alpha(\mathbf{k})]^{-1}$ and $F_m = \sum_\alpha f_\alpha({\bf k}, m \mathbf{Q}) [i k_0 - E_\alpha(\mathbf{k})]^{-1}$, with explicit expressions for $f_\alpha$, and $g_\alpha$ available in Eqs. ~\eqref{18} and ~\eqref{20}.

We present the phase diagram in Fig.~1 within the parameter space of the interaction strength $V_1$ and the chemical potential $\mu$ (grand-canonical ensemble)~\footnote{The energy difference along the orange boundary is determined with a numerical precision of order $10^{-4}$ while the brown boundary is determined with a numerical precision of order $10^{-3}$}. The first-order transition between the BCS and PDW phases occurs at a critical interaction of approximately $V_1 \sim -6t$~\footnote{The value differs from earlier fixed-density studies~\cite{Loder2010}, which omitted the contribution $\mu N$ in canonical ensemble.}. Our results are consistent with the findings reported in Ref.~\cite{PhysRevB.96.224503}, which properly accounted for these effects within the fixed-density ensemble.

\vskip2mm

\subsection{Superfluid Density} The superfluid density is determined
from the response to an external electromagnetic field. This involves the dynamics of both the quasi-particles~\cite{Halperin1979} and the condensate~\cite{Boyack2017}. As a result $n_s$, consists of two distinct parts: a fermionic contribution from the current-current correlator ($n_0$) and a non-trivial contribution from collective Higgs (amplitude) modes, $n_{\text{col}}$, represented by $n_s\equiv n_0 + n_{\text{col}}$.

\begin{figure}
    \includegraphics[width=1\linewidth]{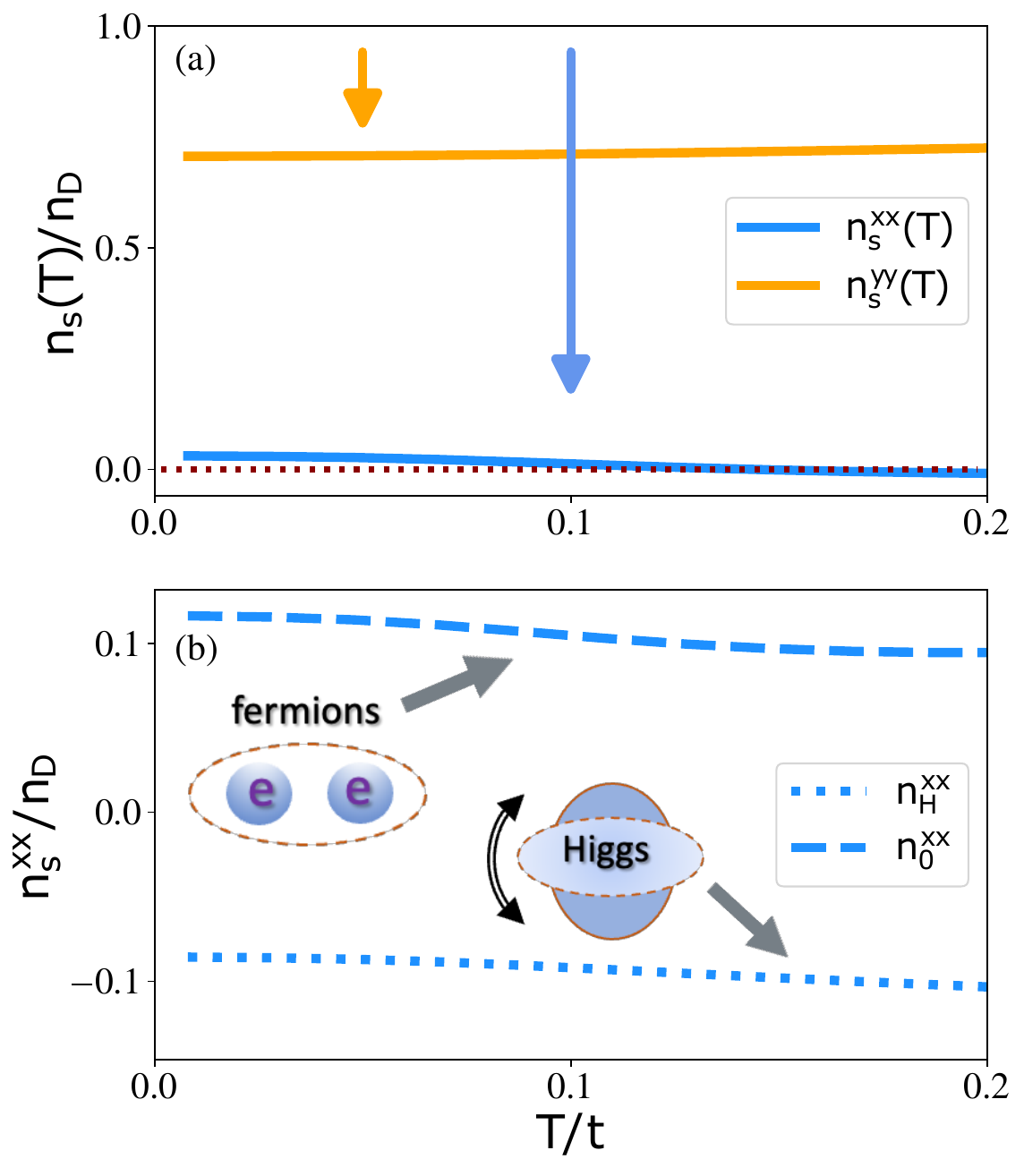}
    \caption{Anisotropic superfluid density in the stable PDW phase for parameters $(-V_1, -\mu) = (5.9t, 0.4t)$. \textbf{(a)} The longitudinal component, $n_s^{xx}$, is dramatically suppressed near zero, in stark contrast to the robust transverse component, $n_s^{yy}$. The superfluid density component $n_s^{xx}$ vanishes at $T_0 \approx 0.15t$, which sets an upper bound for the critical transition temperature $T_c$. Therefore $T_c$ is much smaller than the order parameter at zero temperature.
    \textbf{(b)} This suppression is a two-fold mechanism: first, the fermionic contribution $n_0^{xx}$ is itself suppressed to a small positive value by destructive interference from the large pairing momentum $\mathbf{Q}$. Second, this remaining modest contribution is near-perfectly cancelled by the negative term from the collective Higgs mode, $n_H^{xx}$. This dramatic and fragile anisotropy is a key experimental fingerprint of the PDW state.}
\label{fig2}
\end{figure}

The fermionic contribution to the superfluid density, $n_0^{ii}$, originates from Bogoliubov quasiparticle excitations.We provide a more complete analytical expression in the Methods Section (``Superfluid Density"), 
(which is used to construct Fig.~\ref{fig2}b), but it is instructive to present a more accessible
approximate form which represents the dominant contributions arising from the anomalous correlation functions:
\begin{equation}
n^{ii}_0 \simeq 4 \sum_{\mathbf{p},m \alpha \beta} J_i({\bf p}) J_i({\bf p}'_m)f_\alpha(\mathbf{p},m\mathbf{Q}) f_\beta(\mathbf{p},m\mathbf{Q}) d_{\alpha \beta}(\mathbf{p}).
\label{eq5}
\end{equation}
Here, the $J_i(\mathbf{p}) = \partial_{p_i}\xi$ terms correspond to the fermionic current, evaluated at both the original momentum $\mathbf{p}$ and the momentum shifted by the condensate, $\mathbf{p}'_m = \mathbf{p} - m\mathbf{Q}$ with $m=\pm 1$. The factors $f_\alpha$ are the residues of the anomalous Green's function for each Bogoliubov band $\alpha$, and $d_{\alpha\beta} = [n_F(E_\alpha) - n_F(E_\beta)]/[E_\alpha - E_\beta]$ is a thermal factor representing quasiparticle occupation. The sum over band indices is dominated by a positive, inter-band term integrated over the full Brillouin zone. In contrast, the intra-band term, arising from the gapless Fermi arcs, provides a smaller correction due to its limited phase space.

The collective contribution, $n_{\text{col}}$, arises from amplitude (``Higgs'') fluctuations of the condensate. An external electromagnetic field, $\mathbf{A}$, couples to the finite pair momentum, $\mathbf{Q}$, inducing these fluctuations and generating an additional current response~\cite{Boyack2017, Wang2025PRB}. Our analysis, detailed in the Methods Section (``Collective Modes"), shows that this contribution is always negative. Physically, these collective excitations are energetically costly and therefore must reduce the overall superfluid density.

The Higgs mode contribution, distinguished from Nambu/Goldstone modes, is often overlooked in analyses of the electromagnetic response
\cite{Dai2017}.
However, in PDW phases, the strong suppression of the quasi-particle contribution elevates the relevance of the Higgs mode, 
making the latter essential for a complete characterization of the superfluid response.

\subsection{Destructive Interference}
  The unique properties of the PDW state---its dramatic anisotropy and large region of instability---can be traced to a single underlying mechanism: a {destructive interference effect} originating from the finite momentum $\mathbf{Q}$ of the Cooper pairs. This effect profoundly suppresses the superfluid response along the pairing direction ($\hat{x}$) while leaving the transverse response ($\hat{y}$) intact.

To understand this in more detail, we analyze the expression for the fermionic contribution, Eq.~\eqref{eq5}, by comparing the transverse ($y$) and longitudinal ($x$) responses. The difference in their behavior primarily originates from 
mixed sign contributions in the current-product term, $J_i({\bf p}) J_i({\bf p}'_m)$, 
while the weighting factor $\sum_{\alpha \beta}f_\alpha(\mathbf{p},m\mathbf{Q}) f_\beta(\mathbf{p},m\mathbf{Q}) d_{\alpha \beta}(\mathbf{p})$ is positive. See Eq.~\eqref{26} and ~\eqref{S9}. For the transverse response ($n_s^{yy}$), this term is positive, and interference effects are absent. Although the intra-band contribution through $d_{22}$ is slightly negative, the total response is dominated by the larger, positive inter-band contribution, ensuring the transverse superfluid density is robustly positive.

\begin{figure}
    \centering
    \includegraphics[width=1\linewidth]{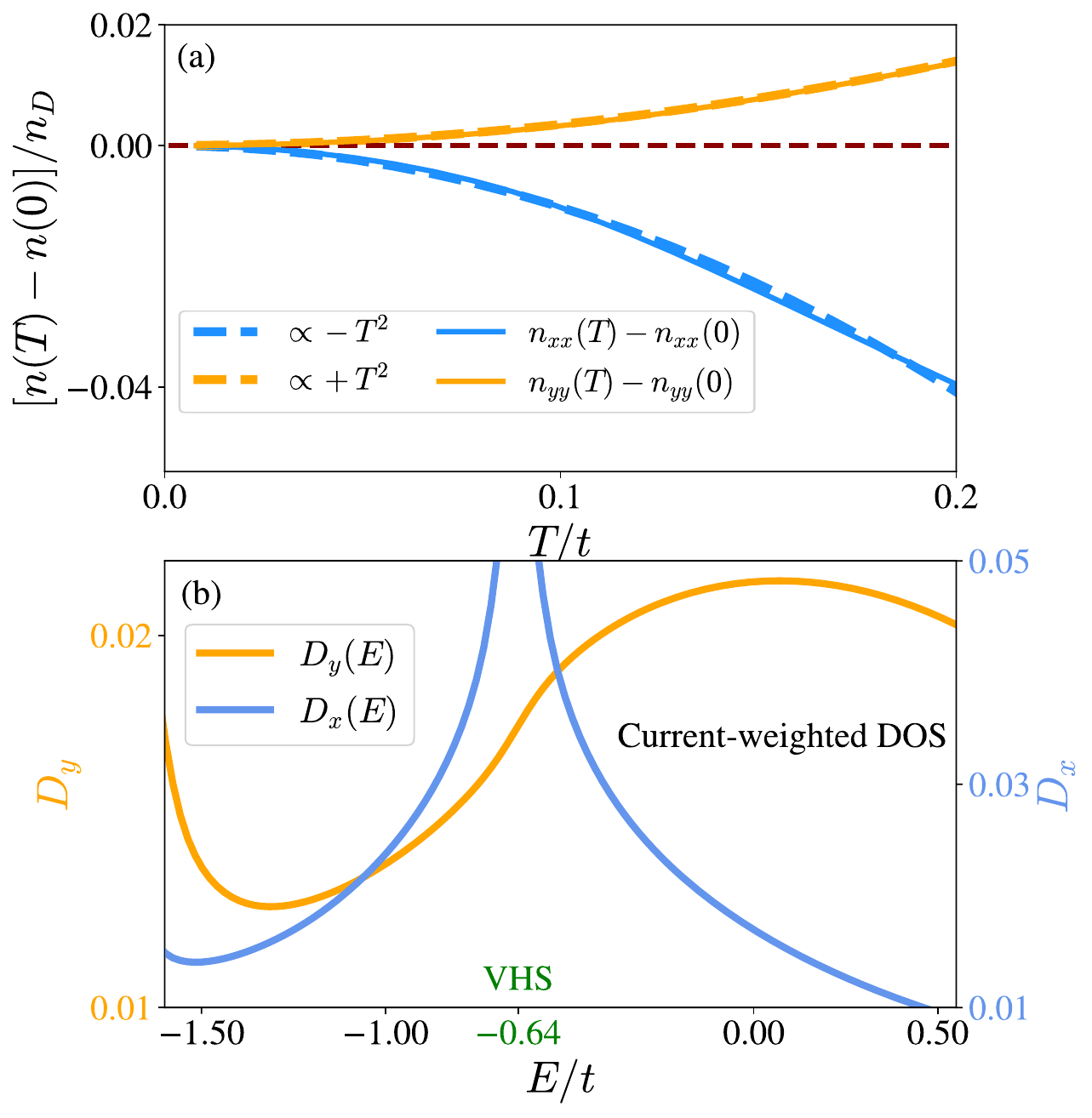}
 \caption{A second key experimental fingerprint of the PDW state: its anomalous temperature dependence, shown here for parameters $(-V_1, -\mu) = (5.8t, 0.5t)$. \textbf{(a)} At low temperatures, the superfluid density exhibits a highly unconventional $T^2$ behavior, increasing along the transverse direction ($n_s^{yy}$) while decreasing along the longitudinal one ($n_s^{xx}$). \textbf{(b)} This striking sign-change is a direct probe of the unique, gapless electronic structure. The opposite signs of the $T^2$ correction are dictated by the opposite curvatures of the longitudinal ($D_x$) and transverse ($D_y$) current-weighted density of states (DOS) at the Fermi level ($E=0$), a feature which arises from a nearby van Hove singularity. For comparison: the homogeneous $s$-wave superfluid density exhibits an exponentially weak $T$-dependence, while the homogeneous $d$-wave shows a linear-in-$T$ decrease at low $T$; both are isotropic with $n_{xx}=n_{yy}$.}
    \label{Superfluid}
\end{figure}

In stark contrast, the longitudinal response ($n_s^{xx}$) is governed by a destructive interference effect. Its mathematical origin is the explicit $Q_x$ dependence within the current-product term:
\begin{eqnarray}
   && J_x\left(\mathbf{p}+\frac{\bf Q}{2}\right)J_x\left(\mathbf{p}-\frac{\bf Q}{2}\right) \nonumber \\
  &&  = 2t^2 (\cos Q_x - \cos p_x) (2t - 4t' \cos p_y)^2. \label{10}
\end{eqnarray}
The $(\cos Q_x - \cos p_x)$ factor is the source of this interference implying that
the product can become negative. 

%
This interference mechanism has two profound consequences. First, it will render the total $n_s^{xx}$ negative for large pairing momenta, leading to the large unstable regime in the phase diagram (Fig.~\ref{phase_diagram}). 
While the precise value of this threshold is model-dependent, the underlying physics of destructive transport 
interference 
arising from 
finite $Q$, as well as a negative contribution from Higgs modes, is general. We therefore expect an upper bound on the pairing momentum to be a universal feature of stable PDW phases.  
Here, specifically, this instability occurs when the pairing momentum exceeds the critical value, $Q_c\simeq 0.44 \pi/a$.

Importantly, even when stable, this interference dramatically suppresses the fermionic contribution $n_0^{xx}$. As shown in Fig.~\ref{fig2}(b), this fermionic term is nearly cancelled by the negative collective Higgs mode
contributions, greatly reducing the total longitudinal superfluid density. Our numerical calculations confirm that this strong suppression is also a robust feature, present across the entire stability region of the PDW phase. 
Moreover, these interference and collective Higgs mode contributions appear generic, that is,
not specific to our model.  

The strong anisotropy in superfluid density could be detected in measurements of optical conductivity. The $f$-sum rule implies that a reduction in superfluid density redistributes spectral weight into 
the finite-frequency $\sigma(\omega)$. Hence, a strongly anisotropic optical response is expected. Concretely, we suggest THz time-domain spectroscopy\cite{PhysRevLett.122.257001} to extract and compare $\sigma_{xx}(\omega)$ and $\sigma_{yy}(\omega)$, where we expect $\sigma_{xx}\gg \sigma_{yy}$.

\subsection{Finite-Temperature Fingerprints} At finite temperatures, we identify a second unambiguous fingerprint of the PDW state which should be experimentally accessible: an anomalous temperature dependence with opposite signs for the longitudinal and transverse superfluid densities, as shown in Fig.~\ref{Superfluid}(a). In stark contrast to typical superconductors, we find that the transverse component, $n_s^{yy}$, appears to \textit{increase} with a $+T^2$ dependence, while the already fragile longitudinal component, $n_s^{xx}$, \textit{decreases} further with a $-T^2$ dependence. This unconventional low temperature behavior is a direct probe of the gapless excitations in the electronic spectrum.

We understand this low-temperature correction to the superfluid density originates dominantly from the gapless bands which, via a Sommerfeld expansion, lead to a $T^2$ dependence:
\begin{equation}
n^{ii }_0(T)-n^{ii}_0(0)\simeq -\frac{\pi^2 T^2}{6} D''_{i}(0) ,\quad \text{for } \frac{T}{\Delta} \ll 1.
\end{equation}
The sign of the temperature correction is therefore determined by the curvature, $D''_{i}(0)$, of the current-weighted density of states (DOS) at the Fermi level, where $D''_{i}(E) \equiv \partial^2 D_i(E) / \partial E^2$. This DOS is formally defined as:
\begin{equation}
D_{i}(E)\equiv \int_{\text{BZ}}\! \frac{\mathrm{d}^2p}{(2\pi)^2}\, \left(\frac{\partial \xi}{\partial p_i}\right)^2 f^2_2(\mathbf{p},\mathbf{Q})  \delta(E-E(\mathbf{p})), \label{dos}
\end{equation}
where the integration spans the Brillouin zone and $E(\mathbf{p})$ is the gapless band dispersion. Due the presence of Bogoliubov Fermi surface, $D_i(E)\neq 0$ and is not simply a monotonic function of $E$. By contrast, in uniform BCS phases, $D_i(E)$ monotonically increases with $E$. This is an essential difference that leads to the novel temperature scaling of the superfluid density associated with the PDW order.

Important to understanding these different temperature dependences is a PDW-induced
van Hove singularity (VHS) located near the Fermi energy and the presence of a Bogoliubov Fermi surface.  
This is a consequence of a rather
unique feature of the quasiparticle band structure.
The position of this VHS relative to the Fermi level determines the sign of $D''_{i}(0)$, leading to opposite curvatures for the two principal directions ($D''_{y}(0) < 0$ and $D''_{x}(0) > 0$) in the stable PDW phase. This directly explains the observed temperature trends. This mechanism is robust: our numerical calculations confirm the VHS exists across a wide range of parameters, and we establish in 
the Methods Section ``Van Hove Singularities" that the singularity itself is a necessary consequence of the large PDW order parameter.

The physics here can be understood more analytically
from the saddle-point nature of the dispersion near the VHS at momentum $\mathbf{P}$, where $E_2(\mathbf{p}+\mathbf{P})-E_2(\mathbf{P}) \propto p_x^2-p_y^2$. This hyperbolic dispersion leads to a density of states (DOS) with a characteristic logarithmic form:
\begin{eqnarray}
  D_{y}(E)\simeq -C\tilde{E}\ln |\tilde{E}|,
\end{eqnarray}
where $\tilde{E}=E-E_2(\mathbf{P})$ and $C$ is a positive constant. The crucial consequence of the relation $D''_{y}(0)\propto 1/E_2(\mathbf{P})$ is that the sign of the DOS curvature is determined by the energy of the singularity itself. This proportionality dictates that a negative curvature ($D''_{y}(0) < 0$) occurs if the VHS lies below the Fermi level ($E_2(\mathbf{P}) < 0$). We find that this is precisely the condition met in the stable PDW phases. This occurs because the VHS energy, $E_2(\mathbf{P})$, is an increasing function of the chemical potential, $\mu$. Since the stable PDW phases are found at low filling (where $\mu$ is negative), the VHS in these states naturally lies below the Fermi level. We emphasize that this VHS is induced by the large PDW order, which is not present in normal states.

In contrast, a similar analysis for the longitudinal ($x$) direction reveals a different functional form for the density of states (DOS), $D_x(E) \propto -\ln |\tilde{E}|$. This leads to a DOS curvature at the Fermi level that is always positive, $D''_x(0) \propto 1/E_2^2(\mathbf{P}) > 0$, which in turn dictates a negative $T^2$ dependence for the superfluid density, $n_s^{xx}(T)$. This decrease with temperature (shown in Fig.~\ref{Superfluid}) would lead to the premature vanishing of the longitudinal superfluid density before the superconducting gap closes, which is potentially problematic.
Together, these opposite signs of the DOS curvature ($D''_x > 0$ and the previously established $D''_y < 0$) provide the microscopic explanation for the anisotropic temperature dependence of $n_s$ shown in Fig.~\ref{Superfluid}(a).


We note that the primary PDW order considered in this manuscript can induce other 
sub-leading types of order, such as a charge density wave. As discussed in the Methods Section (``Secondary Order" and Ref.~\cite{Loder2010}), however, these secondary effects are significantly weaker than the PDW order parameter itself. They, thus, do not influence
the robust features of the superfluid density presented here which can lead to instability---namely their anisotropy and
magnitude suppression.

Possible stabilized phases in the unstable PDW region in the phase diagram are discussed in the Methods section ``Candidates for Unstable PDW". 

Finally, we emphasize the contrast between this work, which treats the PDW as a ground state, and other literature that emphasizes fluctuating PDW order above a bulk superconducting $T_c$~\cite{Lee2014,Setty2023}. A discussion of a fluctuating version of the PDW phase we analyze here, which includes beyond mean field effects is outlined in the Methods Section (``Fluctuation Effects").

\section{Discussion }

In this work, we have presented the fundamental features of the superfluid density in an important
class of quantum materials: namely 2D, unidirectional PDW superconductors where we find serious
problems with stability. In the more limited stable regime,  
we predict (i) an extreme in-plane penetration-depth
anisotropy, (ii) a finite-temperature enhancement of the transverse
stiffness while the longitudinal component is depressed, and (iii) characteristic $T^2$ power laws at low $T$ in both transverse and longitudinal
components of the superfluid density.
Each of these constitutes a falsifiable diagnostic
accessible to experimental probes.
These stability constraints suggest directions for future experimental
investigations in these quantum materials. For instance, penetration-depth measurements
in potential PDW superconductors 
should directly search for the
predicted anisotropy and anomalous power laws.
Among these candidates are
some members of the cuprate family\cite{Du2020} and others possessing the honeycomb structure of the graphene family~\cite{Tsuchiya2016,Roy2010} which might as well include moire materials.

We emphasize that, although it does not represent an exact solution, it is
reasonable to expect the causative mechanisms identified here to be rather generic; they
should be applicable as well to bidirectional PDW superconductors. These involve the strong suppression of superfluid density via destructive interference and the anomalous temperature dependence arising from the interplay of a van Hove singularity and the Bogoliubov Fermi surface. 

Finally, our results serve as a broader cautionary tale for the theoretical study of the larger class of
of novel correlated superconductors. It is essential that theoretical proposals verify the fundamental stability of 
newly discovered phases. 
This task requires a careful calculation of the superfluid density, a non-trivial undertaking that must include the often-overlooked contributions from collective (Higgs) modes. These are not mere technical details; the implications are profound. Without a positive and robust superfluid density, a theoretically proposed phase has no guarantee of being realized in nature.

\section{Methods}

\subsection{Gor'kov Equations} \label{Gorkov}
We begin by defining the propagators
\begin{align}
    G_{ss}({\bf k},{\bf k}',\tau) &= -T_\tau \big\langle c_{{\bf k}s}(\tau)\,c^\dagger_{{\bf k}'s}(0)\big\rangle, \\
    F_{ss'}({\bf k},{\bf k}',\tau) &= T_\tau \big\langle c_{{\bf k}s}(\tau)\,c_{-{\bf k}'s'}(0)\big\rangle.
\end{align}
From their equations of motion one derives the full, non-perturbative Gorkov equations for the unidirectional PDW state:
\begin{align}
        G_{ss}({\bf k},{\bf k}',i\omega_n) &= G_0({\bf k},i\omega_n) \Biggl[ \delta_{{\bf k},{\bf k}'}   - G_0 \sum_m \Delta_{ss'}({\bf k},m{\bf Q})  \nonumber \\
        &\qquad \times F^\dagger_{ss'}\bigl({\bf k}',{\bf k}-m{\bf Q},i\omega_n\bigr) \Biggr], \\
        F_{ss'}({\bf k},{\bf k}',i\omega_n) &= G_0({\bf k},i\omega_n) \sum_{\bf q}\,\Delta_{ss'}({\bf k},{\bf q}) \nonumber \\
        &\quad \times G_{s's'}\bigl(-{\bf k}',-{\bf k}+{\bf q},-i\omega_n\bigr).  
\end{align}
Here $G$ and $F$ are expressed in Matsubara–frequency space.  Note that the normal Green function need not be diagonal in momentum (i.e.\ ${\bf k}\neq{\bf k}'$), reflecting the broken translational invariance due to ${\bf Q}$.  

For practical calculations one combines these two equations and then neglects terms which involve inter-condensate correlations and give only subleading corrections.  Within this approximation\footnote{The approximation has some weakness as it does not preserve the periodicity under ${\bf k}\to{\bf k}+{\bf Q}$.} one finds
\begin{equation}
\begin{split}
    &G({\bf k},{\bf k},i\omega_n) = \\
    &\frac{G_0({\bf k},i\omega_n)}{1 + G_0({\bf k},i\omega_n) \sum_{m} G_0(m{\bf Q}-{\bf k},-i\omega_n) \bigl|\Delta({\bf k},m{\bf Q})\bigr|^2 }.
\end{split}
\end{equation}
Hence $G$ becomes diagonal in ${\bf k}$.  Dropping the residual momentum dependence, one further obtains
\begin{align}
    G({\bf k},i\omega_n) &= \frac{ \prod_{j=1,2} \bigl(i\omega_n + \epsilon_{-{\bf k}+{\bf q}_j}\bigr) }{ \prod_{j=1}^3 \bigl(i\omega_n - E_j\bigr) }, \\
    F_{ss'}({\bf k},i\omega_n) &= \frac{ \Delta_{ss'}({\bf k},m{\bf Q}) \bigl(i\omega_n + \epsilon_{-{\bf k}-m{\bf Q}}\bigr) }{ \prod_{j=1}^3 \bigl(i\omega_n - E_j\bigr) }.
\end{align}
Here $E_1({\bf k})<E_2({\bf k})<E_3({\bf k})$ are the roots of
\begin{equation}
\begin{split}
    0 &= \bigl(x - \epsilon_{\bf k}\bigr) \bigl(x + \epsilon_{-{\bf k} + {\bf Q}}\bigr) \bigl(x + \epsilon_{-{\bf k} - {\bf Q}}\bigr) \\
    &\quad - \sum_{m=\pm}\bigl(x + \epsilon_{-{\bf k} + m{\bf Q}}\bigr) \bigl|\Delta({\bf k},-m{\bf Q})\bigr|^2 .
\end{split}
\end{equation}
We conventionally drop the spin indices in the order parameter, writing $\Delta\equiv\Delta_{\uparrow\downarrow}$.  Equivalently, one can express the propagator in terms of its simple poles,
\begin{align}
  &  G\bigl({\bf k},{\bf k},i\omega_n\bigr) = \sum_{\alpha} \frac{g_\alpha}{i\omega_n - E_\alpha}, \\
  &  g_\alpha({\bf k},{\bf Q}) = \frac{ \bigl(E_\alpha + \epsilon_{-{\bf k}+{\bf Q}}\bigr) \bigl(E_\alpha + \epsilon_{-{\bf k}-{\bf Q}}\bigr) }{ \bigl(E_\alpha - E_\beta\bigr) \bigl(E_\alpha - E_\gamma\bigr) }. \label{18}
\end{align}
and similarly for the anomalous function
\begin{align}
   & F\bigl({\bf k},{\bf k}-{\bf Q},i\omega_n\bigr) = \sum_{\alpha} \frac{f_\alpha}{i\omega_n - E_\alpha}, \\
 &   f_\alpha({\bf k},{\bf Q}) = \frac{ \Delta\bigl({\bf k},{\bf Q}\bigr) \bigl(E_\alpha + \epsilon_{-{\bf k}-{\bf Q}}\bigr) }{ \bigl(E_\alpha - E_\beta\bigr) \bigl(E_\alpha - E_\gamma\bigr) }. \label{20}
\end{align}
where in each case $\alpha,\beta,\gamma$ are all distinct.  From numerical simulations, one can observe that each pole weight $g_\alpha$ (or $f_\alpha$) is nonzero only over a limited region of the Brillouin zone, indicating that $E_\alpha$ describes the excitation energy only in that region.

\begin{figure}
    \centering
    \includegraphics[width=\linewidth]{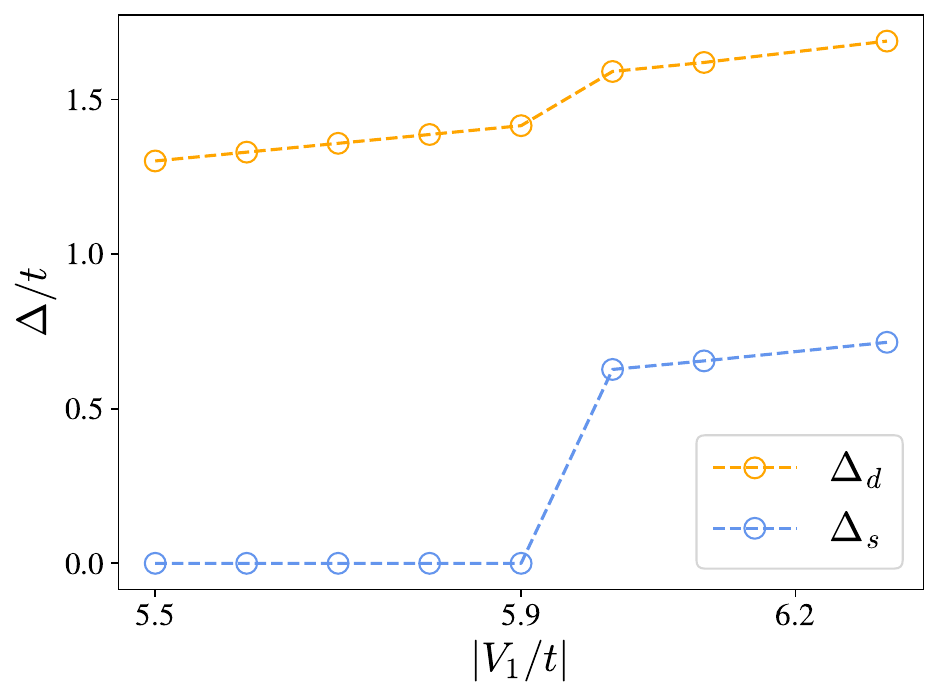}
    \caption{The order parameter is plotted versus the interaction strength $V_1$ at a fixed chemical potential $\mu = -0.5t$. The critical interaction separating the $d$-wave superconducting and Pair-Density Wave (PDW) phases occurs at approximately $V_{1c} = -5.95t$. }
    \label{FS0}
\end{figure}

\subsection{Possible Phases}

We now justify our focus on the competition between uniform $d$-wave BCS superconductivity and a unidirectional PDW state by systematically considering other potential instabilities within our model.

In the absence of interactions, the $t-t'$ model describes a simple metal. When the nearest-neighbor attractive interaction ($V < 0$) is introduced, the system is naturally driven toward instabilities in the particle-particle (superconducting) channel. We can largely rule out other competing orders based on the model's fundamental symmetries and parameter regime. 

The first excluded type is {time-reversal symmetry (TRS) breaking and magnetic phases:} our microscopic model preserves TRS. This makes spontaneous TRS-breaking states energetically unfavorable. For instance, Fulde-Ferrell (FF) states\cite{WangLOFF} ($\Delta(\mathbf{r}) \sim e^{i\mathbf{Q}\cdot\mathbf{r}}$) can be shown to be energetically higher than their TRS-symmetric PDW counterparts\cite{PhysRevB.96.224503}. Likewise, spin-density-wave (SDW) states are disfavored, as they typically require a dominant \textit{repulsive} interaction\cite{PhysRevB.108.064514}, which is absent here.

   The second excluded type is particle-hole channel phases: instabilities in the particle--hole channel, such as a primary charge-density-wave (CDW) order, are also unlikely. A CDW phase usually requires strong, specific Fermi surface nesting, such as $(\pi,\pi)$ nesting at half-filling for $t' = 0$. Our model has $t' \neq 0$, where such a perfect nesting does not occur.

This analysis constrains the primary competition to TRS-symmetric superconducting states. The most natural candidates are: (i) homogeneous $d$-wave BCS, (ii) unidirectional (uni-) PDW, and (iii) bi-directional PDW. While a bi-directional PDW is possible, it generally requires a much stronger attractive interaction to compensate for the additional kinetic energy cost of multiple condensate momenta. Therefore, within the parameter regime studied, the essential physics is captured by the competition between the uniform $d$-wave BCS and uni-PDW phases.

\subsection{Candidates for Unstable PDW}

A key finding of our work (Fig.~1) is the emergence of a uni-PDW phase with a negative superfluid density ($\rho_s < 0$), signaling a thermodynamic instability. This raises the question of the true ground state in this parameter region. We propose two likely candidates for the stable phase that replaces this unphysical solution.

First, for the portion of the unstable regime near the phase boundary $V_{1c}$, we expect the true ground state to be the uniform $d$-wave BCS phase. Our mean-field energy analysis (see Fig.~5) confirms that the BCS solution remains a local energy minimum in this region, with an energy comparable to that of the uni-PDW state. Crucially, the uniform BCS phase is fully gapped and always possesses a positive superfluid density ($\rho_s > 0$), rendering it a thermodynamically stable solution.

Second, deeper within the unstable regime (i.e., further from the $V_{1c}$ boundary), we hypothesize that the system transitions into a {\it coexistence phase} of uniform BCS order and uni-PDW order. Such a mixed state is a natural candidate to resolve the instability. The presence of a uniform BCS component can help gap out the fermionic excitations and mitigate the destructive interference effects, which originate from the pure PDW order. Furthermore, this mixing of orders should be energetically favorable compared to a pure BCS state. A detailed investigation of this potential coexistence phase is a direction for future work.

\subsection{Superfluid Density}
In this section, we explain why PDW superconductors always have a positive transverse superfluid density, while the parallel component can become negative. We begin with the expression for the fermionic contribution to the response function, $K_0^{ij}(k)$:
\begin{equation}
    K_0^{ij}(k) = - \langle \hat{j}_i(k) \hat{j}_j(-k) \rangle + \sum_{\mathbf{p}} \frac{\partial^2 \xi}{\partial p_i \partial p_j} n_\mathbf{p}.
    \label{5}
\end{equation}
Here, the current operator is $\hat{j}_i(k) = \sum_{\mathbf{p}, \sigma} (\partial_{p_i} \xi )c^\dagger_{\mathbf{p} - \mathbf{k}/2, \sigma} c_{\mathbf{p} + \mathbf{k}/2, \sigma}$, and the particle occupation number is denoted by $n_\mathbf{p}= \sum_\sigma\langle c^\dagger_{\mathbf{p} \sigma} c_{\mathbf{p} \sigma} \rangle = 2T\sum_{p_0}G(ip_0,\mathbf{p})$.

The fermionic contribution to the superfluid density, $n_0$, is obtained by taking the zero-frequency, long-wavelength limit of $K_0$. This allows $n_0$ to be written in a compact form for both longitudinal and transverse responses, depending on bilinear combinations of the Gor'kov Green's functions.
This leads to a central equation of our work:
\begin{equation}
n^{ij}_0= \sum_{p,m}  J_i({\bf p}) \left[4 J_j({\bf p}'_m)  F^2_m(p) - 2  \frac{\partial \Delta(\mathbf{p},m\mathbf{Q})}{\partial p_j}  G(p) F_m(p) \right],
\label{S8}
\end{equation}
The second term in Eq. \eqref{S8} is significantly smaller than the first and can be dropped. This simplifies the expression for $n_0$, which can now be written in terms of the residues of the Gor'kov $F$ functions:
\begin{equation}
n^{ii}_0 \simeq 4 \sum_{\mathbf{p},m \alpha \beta} J_i(\mathbf{p})\,J_j(\mathbf{p}'_m) f_\alpha(\mathbf{p},m\mathbf{Q}) f_\beta(\mathbf{p},m\mathbf{Q}) d_{\alpha \beta}(\mathbf{p}),
\label{eq:8}
\end{equation}
where the term $d_{\alpha\beta}$ is given by $d_{\alpha\beta} = [n_F(E_\alpha) - n_F(E_\beta)]/[E_\alpha - E_\beta]$. Here, \( E_\alpha(\mathbf{p}) \) is the Bogoliubov band dispersion and \( n_F \) is the Fermi-Dirac distribution. The current operators are explicitly given by
\begin{eqnarray}
    J_x(\mathbf{p}) &=& 2t\sin(p_x) - 4t'\cos(p_y)\sin(p_x), \nonumber \\
    J_y(\mathbf{p}) &=& 2t\sin(p_y) - 4t'\cos(p_x)\sin(p_y).
\end{eqnarray}

Due to the distinct nature of intra- and inter-band contributions, at $T=0$ we split the contribution into two parts:
\begin{align}
 n^{ij}_{\mathrm{intra}} &= -4\sum_m\int\!\frac{d^2\mathbf{p}}{(2\pi)^2}
    J_i(\mathbf{p})\,J_j(\mathbf{p}'_m) 
    \,f^2(\mathbf{p},m\mathbf{Q})\,\delta\bigl(E(\mathbf{p})\bigr), \\
 n^{ij}_{\mathrm{inter}} &= 8\sum_m\int\!\frac{d^2\mathbf{p}}{(2\pi)^2}
     \,\frac{J_i(\mathbf{p})\,J_j(\mathbf{p}'_m)
    \,f_1(\mathbf{p},m\mathbf{Q})\,f_3(\mathbf{p},m\mathbf{Q})}{E_1-E_3}.\label{26}
\end{align}
Contributions from $n^{ij}_{\mathrm{inter}}$ involve the whole Brillouin Zone (BZ),
while contributions from gapless bands only involve the Fermi surface.
Therefore, $n^{ii}_{\mathrm{inter}} \gg n^{ii}_{\text{intra}}$ for any $i$. This can be understood via the
standard s-wave BCS an example: the intraband term vanishes because there is no Bogoliubov Fermi surface, and
the interband term is purely diamagnetic since the paramagnetic part vanishes. In this case, the
interband term alone gives the total superfluid density.

Thus, below we only consider contributions from gapped bands to show
that $n_{yy}$ is always positive while $n_{xx}$ can be negative.
Firstly, one can derive the relation:
\begin{align}
  &  f_1(\mathbf{p},\mathbf{Q})\,f_3(\mathbf{p},\mathbf{Q}) \frac{1}{E_1-E_3}  = \frac{ \Delta(\mathbf{p},\mathbf{Q})^2}{(E_3-E_1)^3} \times u, \nonumber   \\
  &  u  = \frac{(E_1 + \xi_{-\mathbf{p}+\mathbf{Q}})(E_3 + \xi_{-\mathbf{p}+\mathbf{Q}})}{(E_1 - E_2)(E_3 - E_2)}. \label{S9} 
\end{align}
One can show that $u$ is a positive factor since $E_3>E_2,0>E_1$ and $E_1<-\xi_{-\mathbf{p} + \mathbf{Q}}$ and $E_3>-\xi_{-\mathbf{p} + \mathbf{Q}}$. This justifies the statement in the maintext that the weighting factor is always positive.

Now, let us specifically consider $n_{yy}$. Since $\bf p$ is the momentum in the integral, we can shift the integral variable $\bf  p$ by ${\bf  Q}/2$ and thus consider the product of current operators:
\begin{eqnarray}
&& J_y\left(\mathbf{p}+\frac{\mathbf{Q}}{2}\right)\,J_y\left(\mathbf{p}-\frac{\mathbf{Q}}{2}\right) = 4(t^2-4t'^2)\sin^2\left(Q_x/2\right) \nonumber \\
&&+ \left(4t'\cos p_x - 2t\cos\left(Q_x/2\right)\right)^2  . \nonumber
\end{eqnarray}
 It can be readily observed that if $t \geq t'/2$, then $J_y(\mathbf{p})\,J_y(\mathbf{p}-\mathbf{Q}) \geq 0$, and consequently, $n^{yy}_{0}$ is always positive.
Recall that we are considering the case where $t'/t = 0.3$ and the condition $t \geq t'/2$ is satisfied.
Therefore, we have shown that $n^{yy}_0$ in our case is always positive.

Next, we consider $n_{xx}$ and compute the product of current operators:
\begin{equation}
J_x(\mathbf{p}+ {\mathbf{Q}}/{2})\,J_x(\mathbf{p}- {\mathbf{Q}}/{2}) = 4t^2\frac{\cos Q_x - \cos p_x}{2}\,(2t - 4t'\cos p_y)^2. \label{S10}
\end{equation}
Consider the case where $Q_x=\pi/2$. In this scenario, the expression simplifies to:
\[
J_x(\mathbf{p}+ {\mathbf{Q}}/{2})\,J_x(\mathbf{p}- {\mathbf{Q}}/{2}) = -2t^2\cos p_x\,(2t - 4t'\cos p_y)^2.
\]
Now, we seek numerical insight into why this term leads to a negative contribution. We observe that the peak of $g_2$ is at $k_x \approx 1$, and Eq.~\eqref{S9} assigns more weight to the region $k_x < 1$.
Therefore, $n^{xx}_0$ becomes negative when $Q_x$ is moderately large, say $Q_x=\pi/2$.

\subsection{Van Hove Singularities }
There emerges a new Van Hove singularity (VHS) in the gapless Bogoliubov band. Its position in momentum space differs from that in the free-fermion case. This VHS arises from the order parameter, which couples $k$ with $k\pm Q$. Below, we examine how this VHS originates from the band dispersion.

Let us start from the electronic dispersion $E_\alpha$:
\begin{align}
    E_\alpha(k) &= \frac{\xi_0 - \xi_1 - \xi_2}{3} + \frac{2^{2/3}}{3}\,\Re\!\bigl[\Omega^\alpha\,(A + \mathrm{disc})^{1/3}\bigr], \\
    \Omega &= e^{2\pi i/3}, \nonumber \quad
    \mathrm{disc}  = \sqrt{A^2 + 4\,B^3}. \nonumber
\end{align}
where $\xi_0 = \xi(k)$, $\xi_1 = \xi(k+Q)$, $\xi_2 = \xi(k-Q)$, $\alpha=0,1,2$, $A,B$ are given by
\begin{align}
    A = (2\xi_0 + \xi_1 + \xi_2) \Bigl[ &-2(\xi_1-\xi_2)^2 + \xi_0^2 + \xi_1\xi_2 \\
    &+ \xi_0(\xi_1+\xi_2) + 9\,\Delta^2 \Bigr], \nonumber \\
    B = -(\xi_0 - \xi_1 - \xi_2)^2 &- 3\bigl(2\Delta^2 + \xi_0\xi_1 + \xi_0\xi_2 - \xi_1\xi_2\bigr).
\end{align}
We focus on the gapless band $\alpha=2$. 

First, we observe that these VHS appear in this band over a broad continuous range of parameters $t,t',Q,\mu$.  For concreteness, take the example
\[
t=-\tfrac12,\quad t'=0,\quad Q=\tfrac\pi2,\quad \mu=0.
\]
We recognize that the symmetry factor of the order parameter does not affect the appearance of the VHS, so we treat $\Delta$ as a constant.  Under these conditions, one obtains the identities
\begin{align}
    \xi_1+\xi_2 &= 2\cos k_y,\quad
    \xi_1-\xi_2 = 2\sin k_x, \\
    \xi_1\,\xi_2 &= \cos^2 k_y - \sin^2 k_x.
\end{align}
since $\xi_0 = \cos k_x + \cos k_y,  
    \xi_1 = \sin k_x + \cos k_y,  
    \xi_2 = -\sin k_x + \cos k_y.$
After some algebra, one finds
\begin{align}
    A = 2(\cos k_x + 2\cos k_y) \Bigl[ &(\cos k_x + 2\cos k_y)^2 \\
    &+ 9(\Delta^2 - \sin^2 k_x) \Bigr], \nonumber \\
    B = -\Bigl[ (\cos k_x + 2\cos k_y)^2 &+ 3\sin^2 k_x + 6\Delta^2 \Bigr].
\end{align}

In the large-gap limit $\Delta\gg1$, we have $B^3\sim -\Delta^6$ and $A^2\sim \Delta^4$, so $A^2+4B^3<0$ and we approximate
\begin{equation}
\begin{split}
    (A + \mathrm{disc})^{1/3} \simeq \Delta \Biggl( &18\,\frac{s}{\Delta}\left[1 + \frac{s^2/9 - \sin^2 k_x}{\Delta^2}\right] \\
    &+ i\sqrt{6^3\times4} \Biggr)^{1/3}.
\end{split}
\end{equation}
where $s = \cos k_x + 2\cos k_y$.  Then for the gapless band,
\begin{equation}
\begin{split}
    E_2(k) &\simeq \frac{\xi_0 - \xi_1 - \xi_2}{3} + \frac{2^{2/3}}{3}\, \Re\!\bigl[\Omega^2\,(A + \mathrm{disc})^{1/3}\bigr] \\
    &\simeq \frac{\cos k_x - \cos k_y}{3} - \frac{s}{3}\left[1 + \frac{s^2/9 - \sin^2 k_x}{\Delta^2}\right].
\end{split}
\end{equation}

We locate the VHS by solving $\nabla E_2=0$.  Clearly $\partial E_2/\partial k_y = 0$ at $k_y = \pi$.   For $k_x$, one finds
\begin{equation}
\begin{split}
    \frac{\partial E_2}{\partial k_x} &\simeq \frac{\sin k_x}{3}\, \frac{s^2/3 - \sin^2 k_x + 2s\cos k_x}{\Delta^2} = 0 \\
    &\Longrightarrow k_x \approx 1.505, \text{ when } k_y=\pi.
\end{split}
\end{equation}
in excellent agreement with numerical results.  Thus we have demonstrated the existence of a VHS in this case, whose origin lies in the large $\Delta$ of the order parameter coupling.

\subsection{Collective Mode}
At zero temperature, there are two condensates at $\pm \mathbf{Q}$, which are described by the mean-field solution $\Delta(\pm \mathbf{Q})$. To analyze the collective mode contributions, 
$ K_{\text{col}}$,
we define the operators  
\begin{equation}
     \hat{O}(\mathbf{k}; l m)
  = \sum_{\mathbf{p}} \varphi_l(\mathbf{p})
    \,c_{\mathbf{p}+(m\mathbf{Q}/2 + \mathbf{k}/2) , \uparrow}
    \,c_{-\mathbf{p}+(m\mathbf{Q}/2 + \mathbf{k}/2) , \downarrow}.
\end{equation}
These operators represent the fluctuating (pair) degrees of freedom in the angular momentum basis $l$, expanded around the condensate momenta $\pm \mathbf{Q}$. We treat \(\hat{O}^\dagger\) as an independent variable and thus interpret
\[
  \hat{\Psi}(\mathbf{k}; l m)
  \equiv
  \begin{pmatrix}
    \hat{O}(\mathbf{k}; l m), &
    \hat{O}^\dagger(-\mathbf{k}; l m)
  \end{pmatrix}^T
\]
as the related bosonic degrees of freedom.  We define response functions: $J=\langle \hat{j} \rangle$ and $\Psi=\langle \hat{\Psi} \rangle$. Then one is able to derive the linear response
 \begin{eqnarray}
&& \begin{pmatrix}
{\bf J}\\
{V}_1^{-1}\Psi(q;lm)
\end{pmatrix} \nonumber  \\
=&&\sum_{l'm'}
\begin{pmatrix}
K_0 & R^T(-q;l'm')  \\
R(q;lm) & S(q;lm,l'm')
\end{pmatrix}  
\begin{pmatrix}
-{\bf A}\\
\Psi(q;l'm')
\end{pmatrix}. \nonumber
\end{eqnarray}
Here ${\bf J}$ is the current, and ${\bf A}$ is the external gauge vector. Here the $R$ and $Q$ are defined by
\begin{align}
    R_{ij}(k; l m) &\equiv - \langle \hat{\Psi}_i(k; l m) \hat{j}_j(-k) \rangle, \\
    S(q; l m, l' m') &\equiv - \langle \hat{\Psi}(k; l m) \hat{\Psi}^\dagger(k; l' m') \rangle.
\end{align}
Note that $R,S$ is an $8\times 8$ matrix, which includes two condensates, $s/d$-wave, and amplitude/phase fluctuations.
One can solve for $\Psi$ in terms of $A$, yielding $\Psi=(-{V}_1^{-1}+S)^{-1}R{\bf A}$, where the matrix $(-{V}_1^{-1}+S)^{-1}$ defines the propagators of the collective modes. Subsequently, one can write ${\bf J}=-K{\bf A}$, which leads to 
\begin{eqnarray}
    K_{\text{col}}(k) =  - R^\mathrm{T}(-k) \, [-1/V_1+S(k)]^{-1}R(k) .
\end{eqnarray}
The collective contribution to the superfluid density, $n_{\text{col}}$, is given by the expression $n_{\text{col}}=K_{\text{col}}(i k_0=0, \mathbf{k}\rightarrow0)$.

To simplify the calculation of $K_{\text{col}}$, we first write the vector $\Psi$ in terms of $O$ as follows:
\begin{eqnarray}
    \Psi=[{O}(s+), {O}(d+),{O}(s-), {O}(d-), \nonumber \\ {O}^\dagger (s+), {O}^\dagger (d+),{O}^\dagger (s-), {O}^\dagger (d-)]
\end{eqnarray}
 
Here, $s/d$ denotes the angular momentum $l$, and $\pm$ denotes the condensate index $m=\pm 1$. Based on symmetries, we can simplify the long-wave length limit $S(q\rightarrow 0)$ and $R(q\rightarrow 0)$ to:
\begin{align}
    S(q\rightarrow 0) &= \begin{pmatrix} X & 0 & U & Y \\ 0 & X & Y & U \\ U & Y & X & 0 \\ Y & U & 0 & X \end{pmatrix}, \\
    R(q\rightarrow 0) &= \begin{pmatrix} v & -v & v & -v \end{pmatrix}.
\end{align}
Here, $X, U,$ and $Y$ are $2\times 2$ matrices, and $v$ is a $2$-component vector. After some algebra, the contribution from the collective modes can be written compactly as:
\begin{eqnarray}
\label{46}
K_{\text{col}}=-4 v (U+X-1/V_1-Y)^{-1} v^T .
\end{eqnarray}
Thus, the collective mode contribution is reduced to a quadratic form. We note that the metric matrix $(U+X-1/V_1-Y)$ is always positive definite (its eigenvalues are positive). Therefore, $K_{\text{col}}$ is always negative (if non-zero).

\subsection{Secondary Order}
Pair-density-wave (PDW) order generically induces two secondary forms of order: a uniform “charge-4e” superconducting order and a charge-density-wave (CDW) order. In real space, these can be defined as
\begin{eqnarray}
&&   \Delta_{4e}(\mathbf{r};\,l,\,l') 
\;\equiv\; 
\bigl\langle \hat{O}(\mathbf{r};\,l,+)\,\hat{O}(\mathbf{r};\,l',-)\bigr\rangle,
\\
&& \rho_{\mathrm{CDW}}(\mathbf{r};\,l,\,l') 
\;\equiv\; 
\bigl\langle \hat{O}^\dagger(\mathbf{r};\,l,+)\,\hat{O}(\mathbf{r};\,l',-) 
\;+\;\mathrm{h.c.}\bigr\rangle, \nonumber
\end{eqnarray}
 where $\hat{O}(\mathbf{r};\,l,\pm)$ is the real-space PDW pairing field in the angular momentum basis $l$ and expanded around the condensate momenta.  
 
 Here $\Delta_{4e}(\mathbf{r};\,l,\,l')$ describes a uniform superconducting condensate carrying charge $4e$, since it pairs two PDW components of opposite momentum ($+Q$ and $-Q$). $\rho_{\mathrm{CDW}}(\mathbf{r};\,l,\,l')$ represents a charge-density modulation (CDW) at wavevector $2Q$, because it involves mixing $\pm Q$ PDW components and hence oscillates in space. 

In practice, both $\Delta_{4e}$ and $\rho_{\mathrm{CDW}}$ are subleading compared to the primary PDW order.  Physically, this suppression arises because the two PDW condensates are centered at $\pm Q$, so combining them involves a large momentum transfer $2Q$, which causes destructive interference. Numerically we notice that the charge density order is around one percent of PDW order strength.

These secondary correlations give sub-leading corrections to the collective‐mode spectrum, which are included in the collective response, see Eq.~\eqref{46}. In particular, $\Delta_{4e}(\mathbf{r};\,l,\,l')$ and $\rho_{\mathrm{CDW}}(\mathbf{r};\,l,\,l')$ couple phase and amplitude fluctuations of the two condensates. Consequently, they influence how the PDW’s phase and amplitude modes propagate and mix.

\subsection{Fluctuation Effects}

There is a body of theoretical work~\cite{Lee2014,Setty2023}
which has proposed that in the cuprates the PDW phase exists only
in the normal state, while
below the transition a d-wave, zero momentum paired
superconductor appears. This is a fairly standard scenario in the theoretical literature.
This is in contrast to some cuprate experiments~\cite{Du2020} where such a PDW phase is reported to exist
below $T_c$.
While we have not treated the normal state fluctuation effects in this paper
based on earlier work on Fermi gases~\cite{Chen2007} we have a different perspective about fluctuations from that in Ref.~\onlinecite{Setty2023}
which we present here.

\begin{figure}
    \centering
    \includegraphics[width=\linewidth]{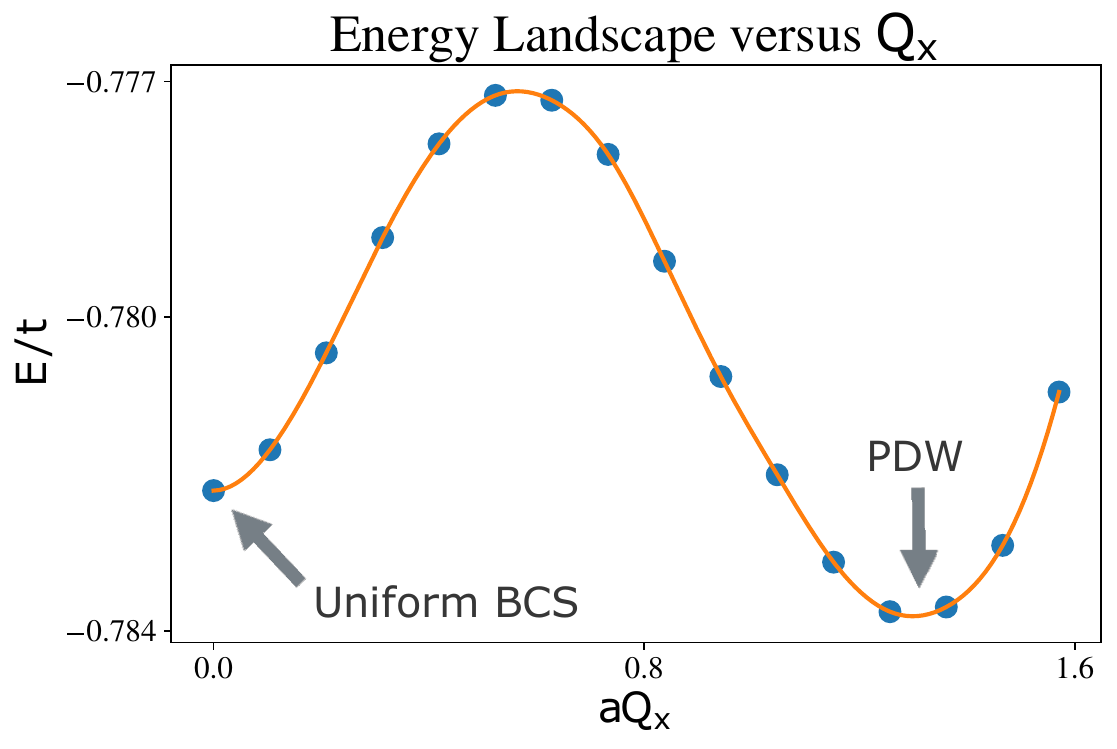}
    \caption{The dependence of the energy of superconducting states on the pairing momentum, $Q_x$, is calculated for a chemical potential of $\mu=-0.5$ and an interaction potential of $V=-6.0$. The results indicate the existence of two local minima, with the global, most stable minimum found at a non-zero value of $Q_x$.}
    \label{FS1}
\end{figure}

Indeed, one can see from Fig.~\ref{FS1} which plots the ground state energy landscape that there are
two meta-stable states for a particular set of parameters in the Hamiltonian, with one energy minimum at $Q_x = 0$ corresponding to a BCS state
and the other at $Q_x \sim 1.3/a$, corresponding to the PDW phase. It is this second minimum that would
be relevant to a fluctuating PDW state.
Moreover, we
argue that
any treatment of PDW fluctuations here
should necessarily be compatible with a state in which there are Fermi arc (gapless) excitations,
as well as two other Bogoliubov bands. Such a state is associated with the PDW not the BCS minimum.
This involves the pair susceptibility~\cite{Chen2007} which is different from that discussed in
Ref.~\onlinecite{Setty2023}
and which is given by
\begin{eqnarray}
\chi_{ll'}(P) 
&=& \sum_{K}\Bigl[\,G_{0\uparrow}(P-K)\,G_{\downarrow}(K)
    + G_{0\downarrow}(P-K)\,G_{\uparrow}(K)\Bigr] \nonumber  \\
 &&   \times \frac{1}{2} \varphi_l\left({\bf k}- {{\bf p}}/{2}\right) \varphi_{l'}\left({\bf k}- {{\bf p}}/{2}\right)  \label{eq:LOFF2symchi} .  
 \end{eqnarray}
Its explicit formula is given by:
 \begin{eqnarray}
\chi_{ll'}(P) 
&=& T\sum_{n}\sum_{\mathbf{k},\sigma}
    \frac{\varphi_l\left({\bf k}- {{\bf p}}/{2}\right) \varphi_{l'}\left({\bf k}- {{\bf p}}/{2}\right)}{\,i\Omega - i\omega_{n} - \xi_{\mathbf{p-k} }\,}\, \nonumber \\
    &&
 \times   \frac{\bigl(i\omega_{n} + \xi_{\mathbf{-k+Q} }\bigr)\,
          \bigl(i\omega_{n} + \xi_{\mathbf{-k-Q} }\bigr)}
         {\,(i\omega_{n} - E_{1 })\,
           (i\omega_{n} - E_{2 })\,
           (i\omega_{n} - E_{3 })\,}  .
\end{eqnarray}
Here, we weight $G_0G$
by symmetry factors because two angular momentum components, $l=s,d$, are involved in PDW situations. When $P=(0,{\bf Q})$, one can write the $\chi(P)$ as 
\begin{eqnarray}
\chi_{ll'}(0,{\bf Q})&=&-T\sum_{n,\mathbf{k},\sigma}
    \frac{\bigl(i\omega_{n} + \xi_{\mathbf{-k-Q} }\bigr)\,
           }
         {\,(i\omega_{n} - E_{1 })\,
           (i\omega_{n} - E_{2 })\,
           (i\omega_{n} - E_{3 })\,}\nonumber \\
      &\times &     \varphi_l({\bf k}-{\bf Q}/2)\varphi_{l'}\left({\bf k}- {{\bf Q}}/{2}\right).
\end{eqnarray}
This pair susceptibility can be used to derive the gap equations for the PDW phase,
\begin{equation}
    \Delta_l(m{\bf Q}) = V_1 \sum_k \varphi_l(\mathbf{k}-m\mathbf{Q}/2) F_m(k)
\end{equation}
This can be rewritten in the form
\begin{equation}
\Delta_l({\bf Q}) = -V_1 \sum_{l'} \chi_{ll'}(0,{\bf Q}) \Delta_{l'}({\bf Q}),
\end{equation} which is equivalent to the matrix equation:
\(
(I+V_1\chi) \Delta({\bf Q})=0
\).
Here, $I$ is the $2\times 2$ {identity} matrix and $\Delta({\bf Q})$ is a two-component vector. This relationship is equivalent to {saying} that the kernel of the matrix $I+V_1\chi$ is not empty, which means its determinant must be zero:
\begin{equation}
   \det (I+V_1\chi) =0.
\end{equation}

To the extent that the literature focus on a PDW addresses the normal state,
one need not be concerned
about the anomalous features of the superfluid density in the PDW phase discussed in this paper.
But the above figure emphasizes that
the nature of these fluctuating pairs associated with a PDW state is different from that
associated with a BCS (d-wave) superconductor. If, as is reasonable $T_c$ represents
a temperature at and below which these \textit{same} normal state preformed pairs condense, one might expect
from this standard scenario that there might be first-order like behavior in
which the condensed and non-condensed pairs are not smoothly connected.
It would be worth investigating whether this standard scenario implies 
that in a material like the high $T_c$ cuprates 
the anti-nodal gap evolves continuously or discontinuously
across $T_c$.

\section{Data availability}
The data analyzed in the current study are available at: 
\href{https://github.com/KernelW/arxiv2506.13631}{https://github.com/KernelW/arxiv2506.13631}

\section{Code availability}
The codes used for the current study are available from the author Ke Wang on reasonable request.

\section{Acknowledgements} We thank Maxim A. Metlitski, Boris Svistunov, and Yiming Wu for helpful discussions. 
Q. C. is supported by the Innovation Program for Quantum Science and Technology (Grant No. 2021ZD0301904).
R.~B. is supported by the Department of Physics and Astronomy, Dartmouth College.
We also acknowledge the University of Chicago's Research Computing Center for their support of this work.

\section{Author Contributions}
K.L. conceived and supervised the project. K.W. performed the computations. K.W. contributed to the acquisition of the data and preparation of figures. All authors have contributed to the interpretation of the numerical data
and the drafting as well as the revision of the manuscript. R.B.
clarified issues involving gauge invariance compatibility in PDW phases
and Q.C. to understanding the behavior of fluctuation effects in PDW
phases.

\section{Competing Interests}
The authors declare no competing interests.

\section{ ADDITIONAL INFORMATION}
Correspondence and requests for materials should be addressed to the authors K. Wang and K. Levin.

\bibliography{ref}

@misc{WangKekule2025,
      title={Kekul\'e Superconductivity in Twisted Magic Angle Bilayer Graphene}, 
      author={Ke Wang and K. Levin},
      year={2025},
      eprint={2510.06451},
      archivePrefix={arXiv},
      primaryClass={cond-mat.supr-con},
      url={https://arxiv.org/abs/2510.06451}, 
}

@Article{Huang2022,
author={Huang, Kevin S.
and Han, Zhaoyu
and Kivelson, Steven A.
and Yao, Hong},
title={Pair-density-wave in the strong coupling limit of the Holstein-Hubbard model},
journal={npj Quantum Materials},
year={2022},
month={Feb},
day={03},
volume={7},
number={1},
pages={17},
issn={2397-4648},
doi={10.1038/s41535-022-00426-w},
url={https://doi.org/10.1038/s41535-022-00426-w}
}

@Article{WuSteven2025,
author={Wu, Yi-Ming
and Chubukov, Andrey V.
and Wang, Yuxuan
and Kivelson, Steven A.},
title={Time-reversal symmetry breaking, collective modes, and Raman spectrum in pair-density-wave states},
journal={npj Quantum Materials},
year={2025},
month={Aug},
day={01},
volume={10},
number={1},
pages={84},
issn={2397-4648},
doi={10.1038/s41535-025-00808-w},
url={https://doi.org/10.1038/s41535-025-00808-w}
}

@article{PhysRevLett.122.257001,
  title = {Infrared Activation of the Higgs Mode by Supercurrent Injection in Superconducting NbN},
  author = {Nakamura, Sachiko and Iida, Yudai and Murotani, Yuta and Matsunaga, Ryusuke and Terai, Hirotaka and Shimano, Ryo},
  journal = {Phys. Rev. Lett.},
  volume = {122},
  issue = {25},
  pages = {257001},
  numpages = {5},
  year = {2019},
  month = {Jun},
  publisher = {American Physical Society},
  doi = {10.1103/PhysRevLett.122.257001},
  url = {https://link.aps.org/doi/10.1103/PhysRevLett.122.257001}
}

@article{PhysRevB.108.064514,
  title = {Superconducting phases of the square-lattice extended Hubbard model},
  author = {Chen, Wei-Chih and Wang, Yao and Chen, Cheng-Chien},
  journal = {Phys. Rev. B},
  volume = {108},
  issue = {6},
  pages = {064514},
  numpages = {13},
  year = {2023},
  month = {Aug},
  publisher = {American Physical Society},
  doi = {10.1103/PhysRevB.108.064514},
  url = {https://link.aps.org/doi/10.1103/PhysRevB.108.064514}
}

@article{PhysRevB.96.224503,
  title = {Effective model for a supercurrent in a pair-density wave},
  author = {W\aa{}rdh, Jonatan and Granath, Mats},
  journal = {Phys. Rev. B},
  volume = {96},
  issue = {22},
  pages = {224503},
  numpages = {11},
  year = {2017},
  month = {Dec},
  publisher = {American Physical Society},
  doi = {10.1103/PhysRevB.96.224503},
  url = {https://link.aps.org/doi/10.1103/PhysRevB.96.224503}
}

@misc{papaj2025,
  title        = {Pair density modulation from glide symmetry breaking and nematic superconductivity},
  author       = {Papaj, Michał and Kong, Lingyuan and Nadj-Perge, Stevan and Lee, Patrick A.},
  year         = {2025},
  eprint       = {2506.19903},
  archivePrefix= {arXiv},
  primaryClass = {cond-mat.supr-con},
  note         = {arXiv:2506.19903}
}

@article{PhysRevLett.127.197003,
  title = {Phonon-Mediated Long-Range Attractive Interaction in One-Dimensional Cuprates},
  author = {Wang, Yao and Chen, Zhuoyu and Shi, Tao and Moritz, Brian and Shen, Zhi-Xun and Devereaux, Thomas P.},
  journal = {Phys. Rev. Lett.},
  volume = {127},
  issue = {19},
  pages = {197003},
  numpages = {7},
  year = {2021},
  month = {Nov},
  publisher = {American Physical Society},
  doi = {10.1103/PhysRevLett.127.197003},
  url = {https://link.aps.org/doi/10.1103/PhysRevLett.127.197003}
}

@Article{Chen2021,
author={Chen, Zhuoyu
and Wang, Yao
and Rebec, Slavko N.
and Jia, Tao
and Hashimoto, Makoto
and Lu, Donghui
and Moritz, Brian
and Moore, Robert G.
and Devereaux, Thomas P.
and Shen, Zhi-Xun},
title={Anomalously strong near-neighbor attraction in doped 1D cuprate chains},
journal={Science},
year={2021},
month={Sep},
day={10},
publisher={American Association for the Advancement of Science},
volume={373},
number={6560},
pages={1235-1239},
doi={10.1126/science.abf5174},
url={https://doi.org/10.1126/science.abf5174}
}

@article{Roy2010,
  title={Unconventional superconductivity on honeycomb lattice: Theory of Kekule order parameter},
  author={Roy, Bitan and Herbut, Igor F},
  journal={Physical Review B—Condensed Matter and Materials Physics},
  volume={82},
  number={3},
  pages={035429},
  year={2010},
  publisher={APS}
}

@article{Tsuchiya2016,
  title={Cooperon condensation and intravalley pairing states in honeycomb Dirac systems},
  author={Tsuchiya, Shunji and Goryo, Jun and Arahata, Emiko and Sigrist, Manfred},
  journal={Physical Review B},
  volume={94},
  number={10},
  pages={104508},
  year={2016},
  publisher={APS}
}

@article{Fulde1964,
  title={Superconductivity in a strong spin-exchange field},
  author={Fulde, Peter and Ferrell, Richard A},
  journal={Physical Review},
  volume={135},
  number={3A},
  pages={A550},
  year={1964},
  publisher={APS}
}

@article{Yin2014,
  title={Fulde-Ferrell states and Berezinskii-Kosterlitz-Thouless phase transition in two-dimensional imbalanced Fermi gases},
  author={Yin, Shaoyu and Martikainen, J-P and T{\"o}rm{\"a}, P{\"a}ivi},
  journal={Physical Review B},
  volume={89},
  number={1},
  pages={014507},
  year={2014},
  publisher={APS}
}

@article{PhysRevLett.60.2677,
  title = {Superconducting Ground State of Noninteracting Particles Obeying Fractional Statistics},
  author = {Laughlin, R. B.},
  journal = {Phys. Rev. Lett.},
  volume = {60},
  issue = {25},
  pages = {2677--2680},
  numpages = {0},
  year = {1988},
  month = {Jun},
  publisher = {American Physical Society},
  doi = {10.1103/PhysRevLett.60.2677},
  url = {https://link.aps.org/doi/10.1103/PhysRevLett.60.2677}
}

@article{PhysRev.108.1175,
  title = {Theory of Superconductivity},
  author = {Bardeen, J. and Cooper, L. N. and Schrieffer, J. R.},
  journal = {Phys. Rev.},
  volume = {108},
  issue = {5},
  pages = {1175--1204},
  numpages = {0},
  year = {1957},
  month = {Dec},
  publisher = {American Physical Society},
  doi = {10.1103/PhysRev.108.1175},
  url = {https://link.aps.org/doi/10.1103/PhysRev.108.1175}
}

@Article{Kong2025,
author={Kong, Lingyuan
and Papaj, Micha{\l}
and Kim, Hyunjin
and Zhang, Yiran
and Baum, Eli
and Li, Hui
and Watanabe, Kenji
and Taniguchi, Takashi
and Gu, Genda
and Lee, Patrick A.
and Nadj-Perge, Stevan},
title={Cooper-pair density modulation state in an iron-based superconductor},
journal={Nature},
year={2025},
month={Apr},
day={01},
volume={640},
number={8057},
pages={55-61},
issn={1476-4687},
doi={10.1038/s41586-025-08703-x},
url={https://doi.org/10.1038/s41586-025-08703-x}
}

@article{Wu1,
  title = {Pair-Density-Wave and Chiral Superconductivity in Twisted Bilayer Transition Metal Dichalcogenides},
  author = {Wu, Yi-Ming and Wu, Zhengzhi and Yao, Hong},
  journal = {Phys. Rev. Lett.},
  volume = {130},
  issue = {12},
  pages = {126001},
  numpages = {8},
  year = {2023},
  month = {Mar},
  publisher = {American Physical Society},
  doi = {10.1103/PhysRevLett.130.126001},
  url = {https://link.aps.org/doi/10.1103/PhysRevLett.130.126001}
}

@article{Wu2,
  title = {Sublattice interference promotes pair density wave order in kagome metals},
  author = {Wu, Yi-Ming and Thomale, Ronny and Raghu, S.},
  journal = {Phys. Rev. B},
  volume = {108},
  issue = {8},
  pages = {L081117},
  numpages = {6},
  year = {2023},
  month = {Aug},
  publisher = {American Physical Society},
  doi = {10.1103/PhysRevB.108.L081117},
  url = {https://link.aps.org/doi/10.1103/PhysRevB.108.L081117}
}

@Article{Chen2023,
author={Chen, Weipeng
and Huang, Wen},
title={Pair density wave facilitated by Bloch quantum geometry in nearly flat band multiorbital superconductors},
journal={Science China Physics, Mechanics {\&} Astronomy},
year={2023},
month={Jul},
day={07},
volume={66},
number={8},
pages={287212},
issn={1869-1927},
doi={10.1007/s11433-023-2122-4},
url={https://doi.org/10.1007/s11433-023-2122-4}
}

@article{Agterberg2020,
  title={The physics of pair-density waves: cuprate superconductors and beyond},
  author={Agterberg, Daniel F and Davis, JC S{\'e}amus and Edkins, Stephen D and Fradkin, Eduardo and Van Harlingen, Dale J and Kivelson, Steven A and Lee, Patrick A and Radzihovsky, Leo and Tranquada, John M and Wang, Yuxuan},
  journal={Annual Review of Condensed Matter Physics},
  volume={11},
  number={1},
  pages={231--270},
  year={2020},
  publisher={Annual Reviews}
}

@article{Shaffer2023,
  title={Weak-coupling theory of pair density wave instabilities in transition metal dichalcogenides},
  author={Shaffer, Daniel and Burnell, FJ and Fernandes, Rafael M},
  journal={Physical Review B},
  volume={107},
  number={22},
  pages={224516},
  year={2023},
  publisher={APS}
}

@article{Rosales2024,
  title={Electronic structure of topological defects in the pair density wave superconductor},
  author={Rosales, Marcus and Fradkin, Eduardo},
  journal={Physical Review B},
  volume={110},
  number={21},
  pages={214508},
  year={2024},
  publisher={APS}
}

@article{Venderley2019,
  title={Evidence of pair-density wave in spin-valley locked systems},
  author={Venderley, Jordan and Kim, Eun-Ah},
  journal={Science advances},
  volume={5},
  number={3},
  pages={eaat4698},
  year={2019},
  publisher={American Association for the Advancement of Science}
}

@article{Soto2017,
  title={Higgs modes in the pair density wave superconducting state},
  author={Soto-Garrido, Rodrigo and Wang, Yuxuan and Fradkin, Eduardo and Cooper, S Lance},
  journal={Physical Review B},
  volume={95},
  number={21},
  pages={214502},
  year={2017},
  publisher={APS}
}

@article{Slagle2020,
  title={Charge transfer excitations, pair density waves, and superconductivity in moir{\'e} materials},
  author={Slagle, Kevin and Fu, Liang},
  journal={Physical Review B},
  volume={102},
  number={23},
  pages={235423},
  year={2020},
  publisher={APS}
}

@article{Yoshida2012,
  title={Pair-density wave states through spin-orbit coupling in multilayer superconductors},
  author={Yoshida, Tomohiro and Sigrist, Manfred and Yanase, Youichi},
  journal={Physical Review B—Condensed Matter and Materials Physics},
  volume={86},
  number={13},
  pages={134514},
  year={2012},
  publisher={APS}
}

@article{Wang2018,
  title={Pair density waves in superconducting vortex halos},
  author={Wang, Yuxuan and Edkins, Stephen D and Hamidian, Mohammad H and Davis, JC S{\'e}amus and Fradkin, Eduardo and Kivelson, Steven A},
  journal={Physical Review B},
  volume={97},
  number={17},
  pages={174510},
  year={2018},
  publisher={APS}
}

@article{Norman2018,
  title={Quantum oscillations in a biaxial pair density wave state},
  author={Norman, MR and Davis, JC S{\'e}amus},
  journal={Proceedings of the National Academy of Sciences},
  volume={115},
  number={21},
  pages={5389--5391},
  year={2018},
  publisher={National Academy of Sciences}
}

@article{Chou2025,
  title={Intravalley spin-polarized superconductivity in rhombohedral tetralayer graphene},
  author={Chou, Yang-Zhi and Zhu, Jihang and Das Sarma, Sankar},
  journal={Physical Review B},
  volume={111},
  number={17},
  pages={174523},
  year={2025},
  publisher={APS}
}

@article{Han2022,
  title={Pair density wave and reentrant superconducting tendencies originating from valley polarization},
  author={Han, Zhaoyu and Kivelson, Steven A},
  journal={Physical Review B},
  volume={105},
  number={10},
  pages={L100509},
  year={2022},
  publisher={APS}
}

@article{Dai2018,
  title={Pair-density waves, charge-density waves, and vortices in high-${T}_c$ cuprates},
  author={Dai, Zhehao and Zhang, Ya-Hui and Senthil, T and Lee, Patrick A},
  journal={Physical Review B},
  volume={97},
  number={17},
  pages={174511},
  year={2018},
  publisher={APS}
}

@article{Dai2017,
  title={Optical conductivity from pair density waves},
  author={Dai, Zhehao and Lee, Patrick A},
  journal={Physical Review B},
  volume={95},
  number={1},
  pages={014506},
  year={2017},
  publisher={APS}
}

@article{Setty2023,
  title={Mechanism for fluctuating pair density wave},
  author={Setty, Chandan and Fanfarillo, Laura and Hirschfeld, PJ},
  journal={Nature Communications},
  volume={14},
  number={1},
  pages={3181},
  year={2023},
  publisher={Nature Publishing Group UK London}
}

@article{Chen2006,
  title={Stability conditions and phase diagrams for two-component {Fermi} gases with population imbalance},
  author={Chen, Qijin and He, Yan and Chien, Chih-Chun and Levin, K},
  journal={Physical Review A—Atomic, Molecular, and Optical Physics},
  volume={74},
  number={6},
  pages={063603},
  year={2006},
  publisher={APS}
}

@article{prokof2000,
  title={Two definitions of superfluid density},
  author={Prokof’ev, Nikolai V and Svistunov, Boris V},
  journal={Physical Review B},
  volume={61},
  number={17},
  pages={11282},
  year={2000},
  publisher={APS}
}

@article{Gubankova2006,
  title={Gapless surfaces in anisotropic superfluids},
  author={Gubankova, E and Mishchenko, EG and Wilczek, F},
  journal={Physical Review B—Condensed Matter and Materials Physics},
  volume={74},
  number={18},
  pages={184516},
  year={2006},
  publisher={APS}
}

@article{Pao2006,
  title={Superfluid stability in the BEC-BCS crossover},
  author={Pao, C-H and Wu, Shin-Tza and Yip, S-K},
  journal={Physical Review B—Condensed Matter and Materials Physics},
  volume={73},
  number={13},
  pages={132506},
  year={2006},
  publisher={APS}
}

@article{Halperin1979,
  title={Resistive transition in superconducting films},
  author={Halperin, BI and Nelson, David R},
  journal={Journal of Low Temperature Physics},
  volume={36},
  pages={599--616},
  year={1979},
  publisher={Springer}
}

@article{Agterberg2008,
  title={Dislocations and vortices in pair-density-wave superconductors},
  author={Agterberg, DF and Tsunetsugu, H},
  journal={Nature Physics},
  volume={4},
  number={8},
  pages={639--642},
  year={2008},
  publisher={Nature Publishing Group UK London}
}

@article{Barci2011,
  title={Role of nematic fluctuations in the thermal melting of pair-density-wave phases in two-dimensional superconductors},
  author={Barci, Daniel G and Fradkin, Eduardo},
  journal={Physical Review B—Condensed Matter and Materials Physics},
  volume={83},
  number={10},
  pages={100509},
  year={2011},
  publisher={APS}
}

@article{Berg2009,
  title={Charge-4e superconductivity from pair-density-wave order in certain high-temperature superconductors},
  author={Berg, Erez and Fradkin, Eduardo and Kivelson, Steven A},
  journal={Nature Physics},
  volume={5},
  number={11},
  pages={830--833},
  year={2009},
  publisher={Nature Publishing Group UK London}
}

@article{Chen2007,
  title={Theory of superfluids with population imbalance: {Finite}-temperature and {BCS-BEC} crossover effects},
  author={Chen, Qijin and He, Yan and Chien, Chih-Chun and Levin, K},
  journal={Physical Review B—Condensed Matter and Materials Physics},
  volume={75},
  number={1},
  pages={014521},
  year={2007},
  publisher={APS}
}

@article{Boyack2017,
  title = {Collective mode contributions to the Meissner effect: {Fulde-Ferrell} and pair-density wave superfluids},
  author = {Boyack, Rufus and Wu, Chien-Te and Anderson, Brandon M. and Levin, K.},
  journal = {Phys. Rev. B},
  volume = {95},
  issue = {21},
  pages = {214501},
  numpages = {6},
  year = {2017},
  month = {Jun},
  publisher = {American Physical Society},
  doi = {10.1103/PhysRevB.95.214501},
  url = {https://link.aps.org/doi/10.1103/PhysRevB.95.214501}
}

@article{Wang2025PRB,
  title = {Higgs amplitude mode in optical conductivity in the presence of a supercurrent: Gauge-invariant formulation with disorder},
  author = {Wang, Ke and Boyack, Rufus and Levin, K.},
  journal = {Phys. Rev. B},
  volume = {111},
  issue = {14},
  pages = {144512},
  numpages = {8},
  year = {2025},
  month = {Apr},
  publisher = {American Physical Society},
  doi = {10.1103/PhysRevB.111.144512},
  url = {https://link.aps.org/doi/10.1103/PhysRevB.111.144512}
}

@article{PRL1977K,
  title = {Universal Jump in the Superfluid Density of Two-Dimensional Superfluids},
  author = {Nelson, David R. and Kosterlitz, J. M.},
  journal = {Phys. Rev. Lett.},
  volume = {39},
  issue = {19},
  pages = {1201--1205},
  numpages = {0},
  year = {1977},
  month = {Nov},
  publisher = {American Physical Society},
  doi = {10.1103/PhysRevLett.39.1201},
  url = {https://link.aps.org/doi/10.1103/PhysRevLett.39.1201}
}

@article{Brian2010PRL,
  title = {Entanglement Entropy and the {Fermi} Surface},
  author = {Swingle, Brian},
  journal = {Phys. Rev. Lett.},
  volume = {105},
  issue = {5},
  pages = {050502},
  numpages = {4},
  year = {2010},
  month = {Jul},
  publisher = {American Physical Society},
  doi = {10.1103/PhysRevLett.105.050502},
  url = {https://link.aps.org/doi/10.1103/PhysRevLett.105.050502}
}

@article{Lee2014,
  title = {Amperean Pairing and the Pseudogap Phase of Cuprate Superconductors},
  author = {Lee, Patrick A.},
  journal = {Phys. Rev. X},
  volume = {4},
  issue = {3},
  pages = {031017},
  numpages = {13},
  year = {2014},
  month = {Jul},
  publisher = {American Physical Society},
  doi = {10.1103/PhysRevX.4.031017},
  url = {https://link.aps.org/doi/10.1103/PhysRevX.4.031017}
}

@article{Loder2010,
  title = {Superconducting state with a finite-momentum pairing mechanism in zero external magnetic field},
  author = {Loder, Florian and Kampf, Arno P. and Kopp, Thilo},
  journal = {Phys. Rev. B},
  volume = {81},
  issue = {2},
  pages = {020511},
  numpages = {4},
  year = {2010},
  month = {Jan},
  publisher = {American Physical Society},
  doi = {10.1103/PhysRevB.81.020511},
  url = {https://link.aps.org/doi/10.1103/PhysRevB.81.020511}
}

@Article{Zhao2023,
author={Zhao, He
and Blackwell, Raymond
and Thinel, Morgan
and Handa, Taketo
and Ishida, Shigeyuki
and Zhu, Xiaoyang
and Iyo, Akira
and Eisaki, Hiroshi
and Pasupathy, Abhay N.
and Fujita, Kazuhiro},
title={Smectic pair-density-wave order in EuRbFe4As4},
journal={Nature},
year={2023},
month={Jun},
day={01},
volume={618},
number={7967},
pages={940-945},
issn={1476-4687},
doi={10.1038/s41586-023-06103-7},
url={https://doi.org/10.1038/s41586-023-06103-7}
}

@Article{Du2020,
author={Du, Zengyi
and Li, Hui
and Joo, Sang Hyun
and Donoway, Elizabeth P.
and Lee, Jinho
and Davis, J. C. S{\'e}amus
and Gu, Genda
and Johnson, Peter D.
and Fujita, Kazuhiro},
title={Imaging the energy gap modulations of the cuprate pair-density-wave state},
journal={Nature},
year={2020},
month={Apr},
day={01},
volume={580},
number={7801},
pages={65-70},
issn={1476-4687},
doi={10.1038/s41586-020-2143-x},
url={https://doi.org/10.1038/s41586-020-2143-x}
}

@article{WangLOFF,
   author = {Jibiao Wang and Yanming Che and Leifeng Zhang and Qijin Chen},
   doi = {10.1103/PhysRevB.97.134513},
   issn = {2469-9950},
   issue = {13},
   journal = {Physical Review B},
   month = {4},
   pages = {134513},
   publisher = {American Physical Society},
   title = {Instability of Fulde-Ferrell-Larkin-Ovchinnikov states in atomic Fermi gases in three and two dimensions},
   volume = {97},
   url = {https://link.aps.org/doi/10.1103/PhysRevB.97.134513},
   year = {2018}
}

\clearpage

\end{document}